\documentstyle[12pt]{article}
\begin{document}
\begin{titlepage}

\centerline{\large \bf Evolution of twist-3 fragmentation functions in
multicolour QCD}
\centerline{\large \bf and the Gribov-Lipatov reciprocity.}
\vspace{10mm}

\centerline{\bf A.V. Belitsky}
\vspace{10mm}

\centerline{\it Bogoliubov Laboratory of Theoretical Physics}
\vspace{3mm}
\centerline{\it Joint Institute for Nuclear Research}
\vspace{3mm}
\centerline{\it 141980, Dubna, Russia}
\vspace{20mm}

\centerline{\bf Abstract}

\vspace{5mm}

It is shown that twist-3 fragmentation functions in the large-$N_c$
limit of QCD obey the DGLAP evolution equations. The anomalous
dimensions are found explicitly. The Gribov-Lipatov reciprocity is
violated at the twist-3 level in the leading logarithmic
approximation.

\hspace{0.5cm}

\end{titlepage}


{\large \bf 1.} A renewed interest in the high energy spin physics
in the last years has been concentrated on the transverse spin
phenomena in hard processes. It revives many ideas developed over
a decade ago. In particular, the notion of the twist-2 transversity
distribution $h_1(x)$, first mentioned by Ralston and Soper
\cite{ral79} has been reinvented as well as its twist-3 counterparts
have been addressed \cite{art90}. Due to chirality conservation
$h_1(x)$ cannot appear in the inclusive deep inelastic scattering
(DIS) but it can be measured, for instance, in Drell-Yan reactions
through the collision of the transversely polarized hadrons
\cite{ral79,art90} and in semi-inclusive pion production in DIS
on the nucleon \cite{jaf93}. In the last case, it enters into the
cross section as a leading contribution together with the twist-3
chiral-odd spin-independent fragmentation function ${\cal I}(\zeta)$
\cite{jaf93}. It is well known that there is considerable
difference between the structure and fragmentation functions.
Namely, the moments of the former are expressed in terms of reduced
matrix elements of the tower of the local operators of definite
twist. This property is established by exploiting the Wilson
operator product expansion for inclusive DIS. Although the
light-cone expansion for the fragmentation processes
is similar to DIS, the moments of the corresponding functions are
not related to any short-distance limit. As a substitute for
the local operators come the Mueller's time-like cut vertices
\cite{muel78} which are essentially nonlocal in the coordinate space
so that the analogy to the operator language is only useful mnemonic.
Nevertheless, it is possible to give the definition of the twist
for them in a broader sense without appealing to the concept of
local operators. A thorough discussion of this issue can be found
in Ref. \cite{bal91}.

The $Q^2$-dependence of the twist-3 distributions has extensively 
been dis\-cus\-sed in the li\-te\-ra\-ture
\cite{evol-even,evol-odd,ali91,bbkt96,bel97}. As has been first
observed in Ref. \cite{ali91}, in multicolour QCD ($N_c \to \infty$)
the flavour nonsinglet twist-3 part of the structure function
$g_2(x)$ obeys a simple DGLAP evolution equation, as in the case of
the twist-2 operators, and the corresponding anomalous dimensions
are known analytically. The same pattern is followed by the
chiral-odd distributions \cite{bbkt96,bel97}. Since the twist-3
fragmentation functions enter into several cross sections on the
same footing as the distributions, their scale dependence is of
great interest. Apart from significance for phenomenology, it is
important for theoretical reasons: while it is know that in the
leading order of the coupling constant the splitting functions for
the twist-2 fragmentation functions can be found from the
corresponding space-like quantities via the Gribov-Lipatov
reciprocity relation \cite{grlip72_2}, no such equality is known
for higher twists.


{\large \bf 2.} As distinguished from the leading twist evolution,
the twist-3 two-quark fragmentation functions receive contribution
from the quark-gluon correlators even in the limit of 
asympto\-ti\-cal\-ly large momentum transfer. To solve the problem, 
one should correctly account for the mixing of correlators of the same 
twist and quantum numbers in the course of renormalization. Recently, 
we have addressed ourselves \cite{bel96} to the question of the
$Q^2$-dependence of the non\-po\-la\-rized chiral-odd (NCO) 
fragmentation function ${\cal I}(\zeta)$ which is given in QCD by the 
Fourier transform along the null-plane of the appropriate matrix 
element of the parton field correlator \cite{col82}
\begin{equation}
{\cal I}(\zeta)
=\frac{1}{4}\int \frac{d\lambda}{2\pi}
e^{i\lambda\zeta}
\langle 0 |
\psi (\lambda n) |h,X \rangle
\langle h,X|\bar \psi (0) |0 \rangle.
\end{equation}
The summation over $X$ is implicit and covers all possible hadronic
final states populated by the quark fragmentation. We use the
light-cone gauge $B_+ = 0$, otherwise a link factor should be
inserted in between the quark fields to maintain the gauge
invariance. By exploiting the equation of motion for the Heisenberg
fermion field operator we can express ${\cal I}(\zeta)$ in terms of
the three-parton correlator ${\cal Z}(\zeta',\zeta)$ which
explicitly involves the gluon field, namely (neglecting the quark
mass)
\begin{equation}
\label{eqmot}
{\cal I}(\zeta) = \int d\zeta' {\cal Z}(\zeta',\zeta),
\end{equation}
where we have introduced the $C$-even quantity
\begin{eqnarray}
&&{\cal Z} (\zeta',\zeta)
=\frac{1}{2}
\left[ {\cal Z}^{(1)}(\zeta',\zeta)+
[{\cal Z}^{(1)}(\zeta',\zeta)]^*
\right]
\end{eqnarray}
with
\begin{equation}
{\cal Z}^{(1)}(\zeta',\zeta)
=\frac{1}{4\zeta}\int \frac{d\lambda}{2\pi} \frac{d\mu}{2\pi}
e^{i\lambda\zeta-i\mu\zeta'}
\langle 0 | {\rm g}\gamma_+ \gamma^\perp_\rho
\psi (\lambda n) |h,X \rangle
\langle h,X|\bar \psi (0) B^\perp_\rho (\mu n)|0 \rangle .
\end{equation}
As we have observed in Ref. \cite{bel96}, in multicolour limit,
{\it i.e.} neglecting the terms ${\cal O}(1/N_c^2)$, an additional
three-parton correlator of the type $\langle 0 | \bar\psi \psi
|h,X \rangle \langle h,X| B^\perp |0 \rangle $, which appears
only through the radiative corrections, decouples from the evolution
equation for ${\cal Z}(\zeta',\zeta)$ and the latter becomes
homogeneous provided we discard the quark-mass effects.
Thus, in the large-$N_c$ limit the RG equation takes the form
\begin{equation}
\mu^2 \frac{\partial}{\partial\mu^2} {\cal Z}(\zeta', \zeta)
= \frac{\alpha}{4\pi}
\int dz' \frac{dz}{z} \theta (\zeta - z)
{\cal K} (z, z', \zeta, \zeta') {\cal Z}(z', z)
\end{equation}
and the evolution kernel is given by the following expression:
\begin{eqnarray}
\frac{1}{N_c}{\cal K} (z, z', \zeta, \zeta')
&=& 2 \frac{z}{\zeta} \delta (\zeta' - \zeta + z - z')
- \delta (\zeta' - \zeta + z) \nonumber\\
&&\hspace{-3cm} - \,\frac{2}{\frac{\zeta}{z}
\left( 1 - \frac{\zeta}{z}\right)}
\delta (\zeta' - \zeta + z - z')
+ 2 \int_{1}^{\infty} \frac{dz''}{z''(1-z'')}
\delta \left( 1 - \frac{\zeta}{z} \right)
\delta (\zeta' - z') \nonumber\\
&&\hspace{-3cm} + \,\left[
\delta (\zeta' - \zeta + z - z')
- \delta (\zeta' - \zeta + z)
\right]
\left[
\frac{z}{z'} - \frac{z(z' + \zeta)}{\zeta (z' - z + \zeta)}
\right] \nonumber\\
&&\hspace{-3cm} + \,\delta \left( 1 - \frac{\zeta}{z} \right)
\biggl\{
\frac{3}{2} \delta (\zeta' - z')
- \ln \left( 1 - \frac{z'}{z} \right)\delta (\zeta' - z')
- 2 \left[ \frac{z'}{\zeta' - z'}
\Theta^0_{11} (\zeta', \zeta' - z')\right]_+ \nonumber\\
&&\hspace{-3cm} + \,\frac{z - \zeta'}{\zeta' - z'}
\left[
\frac{\zeta'}{z - z'} + \frac{z'}{\zeta'}
-\frac{\zeta' - z'}{z}
\right]
\Theta^0_{11} (\zeta', \zeta' - z)
+ \left[
\frac{\zeta'}{z - z'} + \frac{z'}{\zeta'} - 1
\right]
\Theta^0_{11} (\zeta', \zeta' - z')
\biggr\}.
\end{eqnarray}
Here we have used the shorthand notation $\Theta^0_{11}$
for the following step function
\begin{equation}
\label{Theta}
\Theta^0_{11} (z, z')
=\frac{\theta(z) -\theta(z')}{z - z'}.
\end{equation}

Inspired by our knowledge acquired from the study of the
twist-3 structure functions \cite{ali91,bbkt96,bel97}, where
the solution of the asymptotic equations was given by the
convolution of the three-particle correlator with a certain weight
function that is essentially the same as entering into the equation
that gives rise to the dynamic twist-3 contribution to the
two-quark distribution at the tree level, we are able to check
that Eq. (\ref{eqmot}) satisfies the ladder-type evolution equation
with the following splitting function:
\begin{equation}
\label{ker}
\frac{1}{N_c}\int d\zeta' {\cal K} (z, z', \zeta, \zeta')
= - \frac{2}{\left[ \frac{\zeta}{z}
\left( 1 - \frac{\zeta}{z}\right) \right]_+}
+ 2 \frac{z}{\zeta} - 1 + \frac{1}{2}
\delta \left( 1 - \frac{\zeta}{z} \right).
\end{equation}
Thus, for the moments we obtain the following solution of the RG
equation ($Q>Q_0$):
\begin{equation}
\label{moments}
\int_{1}^{\infty} \frac{dz}{z^n} {\cal I} (z, Q)
= \left( \frac{\alpha(Q)}{\alpha(Q_0)} \right)
^{^{\rm NCO}\gamma_n/\beta_0}
\int_{1}^{\infty} \frac{dz}{z^n} {\cal I} (z, Q_0),
\end{equation}
and the corresponding anomalous dimensions equal
\begin{equation}
\label{NCO}
{^{\rm NCO}\gamma}_n = N_c
\left\{
- 2 \psi (n-1) - 2 \gamma_E - \frac{3}{n-1} + \frac{1}{2}
\right\},
\end{equation}
as usual $\beta_0 = \frac{2}{3}N_f - \frac{11}{3}C_A$.

As we have previously mentioned, there exists an equation which
states that in the leading log approximation the time-like (TL)
and space-like (SL) kernels corresponding to the twist-2
parton densities are directly related, in the physical regions
of the corresponding channels, by the Gribov-Lipatov equation
$P^{\rm SL}(x) = P^{\rm TL}(1/x)$, or in terms of the corresponding
anomalous dimensions it looks like $\gamma_{n+2}^{\rm TL} = \gamma_n
^{\rm SL}$. Comparing the result given by Eq. (\ref{NCO}) with the
large-$N_c$ anomalous dimensions known in the literature for the
chiral-odd distribution $e(x)$ \cite{bbkt96,bel97} we see
that there is no universality of the corresponding twist-3 evolution
kernels, {\it i.e.} the Gribov-Lipatov reciprocity is violated.

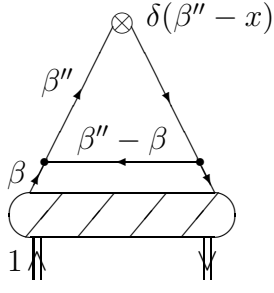
\begin{figure}[t]
\unitlength=2.10pt
\special{em:linewidth 0.4pt}
\linethickness{0.4pt}
\begin{picture}(128.33,51.67)
\
\
\put(106.34,50.19){\line(-1,-2){15.00}}
\put(109.67,50.19){\line(1,-2){15.00}}
\put(108.00,50.86){\makebox(0,0)[cc]{$\otimes$}}
\put(108.00,16.19){\oval(40.67,8.00)[]}
\put(90.67,12.19){\line(5,6){6.67}}
\put(118.67,12.19){\line(5,6){6.67}}
\put(100.01,12.19){\line(5,6){6.67}}
\put(109.34,12.19){\line(5,6){6.67}}
\put(92.00,12.19){\line(0,-1){7.67}}
\put(124.00,12.19){\line(0,-1){7.67}}
\put(122.67,12.19){\line(0,-1){7.67}}
\put(93.34,12.19){\line(0,-1){7.67}}
\put(123.34,7.86){\makebox(0,0)[cc]{$\vee$}}
\put(92.67,7.86){\makebox(0,0)[cc]{$\wedge$}}
\put(94.00,25.67){\line(1,0){28.00}}
\put(94.00,25.67){\circle*{1.33}}
\put(122.00,25.67){\circle*{1.33}}
\put(112.33,51.67){\makebox(0,0)[lc]{$\delta(\beta'' - x)$}}
\put(99.67,36.99){\vector(1,2){1.00}}
\put(115.33,39.00){\vector(1,-2){1.33}}
\put(107.67,25.67){\vector(-1,0){1.00}}
\put(92.00,21.66){\vector(1,2){1.00}}
\put(122.67,24.33){\vector(1,-2){1.33}}
\put(88.67,8.00){\makebox(0,0)[cc]{$1$}}
\put(89.00,23.33){\makebox(0,0)[cc]{$\beta$}}
\put(96.33,40.00){\makebox(0,0)[cc]{$\beta''$}}
\put(108.00,29.67){\makebox(0,0)[cc]{$\beta'' - \beta$}}
\end{picture}
\caption{\label{ladder} One-loop ladder-type diagram for the
two-particle evolution kernel.}
\end{figure}

{\large \bf 3.} Now we can proceed further and demonstrate that
the evolution kernels for the time-like two-quark densities
can directly be found from their space-like analogues by exploiting
the particular form of the evolution equations given in
Ref. \cite{bel97}. Since the analytic structure of the uncut diagram
(see fig.~\ref{ladder}) is completely characterized by the
integral representation of the $\Theta$-function given by
Eq. (\ref{Theta}), we can just take its particular discontinuities,
using the usual Cutkosky rules supplied with appropriate
theta-function specifying the positivity of the energy flow from
the right- to the left-hand side of the cut, in order to obtain the
corresponding time-like kernel. Since the observed particle is
always in the final state for the fragmentation process, we are
restricted to the single cut\footnote{For a given graph the possible
cuts correspond to the possible final states.} across the horizontal
rank of the ladder diagram. Namely, using the integral representation
of the corresponding step function (\ref{Theta}), we have
\begin{equation}
\Theta^0_{11}
(x,x - \beta)=\int_{-\infty}^{\infty}\frac{d\alpha}{2\pi i}
\frac{1}{[\alpha x - 1 +i0][\alpha (x - \beta) - 1 +i0]}
\stackrel{\rm disc}{\longrightarrow}
-\frac{\theta (x - \beta)}{\beta}.
\end{equation}
The self-energy insertions are not affected by the cut since it does
not cross the corresponding lines. Taking into account different
kinematic definitions of the correlation functions in the space-
and time-like regions \cite{col82}, we are able to find the kernels.
It is easy to verify that the evolution kernels constructed for
the time-like twist-2 cut vertices using this recipe coincide
with the known results. In the same way, we may obtain the above
equation (\ref{ker}) from Eq. (90) of Ref. \cite{bel97}.

Since there exist fragmentation functions corresponding to each
distribution, apart from the specific ones appearing from the final
state interaction, we are in a position to find large-$N_c$
anomalous dimensions, which govern their $Q^2$-dependence, from the
results of Refs. \cite{ali91,bbkt96,bel97,mul96}. Namely, the
genuine twist-3 contributions ${\cal H}^{\rm tw-3}_L$ and
${\cal G}^{\rm tw-3}_T$ to the corresponding fragmentation functions
\begin{eqnarray}
&&\hspace{-0.7cm}{\cal H}_L(\zeta)
=\frac{1}{4}\int \frac{d\lambda}{2\pi}
e^{i\lambda\zeta}
\langle 0 |
i \sigma _{+-} \gamma_5 \psi (\lambda n) |h,X \rangle
\langle h,X|\bar \psi (0) |0 \rangle ,\\
&&\hspace{-0.7cm}{\cal G}_T(\zeta)
=\frac{1}{4}\int \frac{d\lambda}{2\pi}
e^{i\lambda\zeta}
\langle 0 |
\gamma_\perp \gamma_5 \psi (\lambda n) |h,X \rangle
\langle h,X|\bar \psi (0) |0 \rangle .
\end{eqnarray}
after subtracting out the twist-2 piece \cite{bal91} obey the
evolution equation (\ref{moments}) with the following anomalous
dimensions:
\begin{eqnarray}
{^{\rm PCO}\gamma}_n &=& N_c
\left\{
- 2 \psi (n-1) - 2 \gamma_E + \frac{1}{n-1} + \frac{1}{2}
\right\},\\
{^{\rm PCE}\gamma}_n &=& N_c
\left\{
- 2 \psi (n-1) - 2 \gamma_E - \frac{1}{n-1} + \frac{1}{2}
\right\}.
\end{eqnarray}
Of course, it is a trivial task to invert the moments and to find
the DGLAP kernels themselves.

To summarize, we have found that in the multicolour limit of QCD
the twist-3 fragmentation functions obey the ladder-type evolution
equations and corresponding anomalous dimensions are known
analytically. This gives a possibility to analyze experimental
data when they become available. The Gribov-Lipatov reciprocity is
not the property of the twist-3 distributions but rather it is
strongly violated already in the leading logarithmic approximation.

This work was supported by Russian Foundation for Fundamental
Research, grant N 96-02-17631.

\end{document}